# OpenFlow Arbitrated Programmable Network Channels for Managing Quantum Metadata

**Venkat R. Dasari[1], and Travis S. Humble[2]**

[1] Army Research Laboratory, 320 Hopkins Road, Aberdeen Proving Ground, Maryland 21004; E-Mails: venkateswara.r.dasari.civ@mail.mil

[2] Quantum Computing Institute, Oak Ridge National Laboratory, One Bethel Valley Road, Oak Ridge, Tennessee 37831; E-Mail: humblets@ornl.gov

\* Author to whom correspondence should be addressed; E-Mail: humblets@ornl.gov; Tel.: +1-865-574-6162; Fax.: +1-865-574-0405

**Abstract:** Quantum networks must classically exchange complex metadata between devices in order to carry out information for protocols such as teleportation, super-dense coding, and quantum key distribution. Demonstrating the integration of these new communication methods with existing network protocols, channels, and data forwarding mechanisms remains an open challenge. Software-defined networking (SDN) offers robust and flexible strategies for managing diverse network devices and uses. We adapt the principles of SDN to the deployment of quantum networks, which are composed from unique devices that operate according to the laws of quantum mechanics. We show how quantum metadata can be managed within a software-defined network using the OpenFlow protocol, and we describe how OpenFlow management of classical optical channels is compatible with emerging quantum communication protocols. We next give an example specification of the metadata needed to manage and control QPHY behavior and we extend the OpenFlow interface to accommodate this quantum metadata. We conclude by discussing near-term experimental efforts that can realize SDN's principles for quantum communication.

This manuscript has been authored by the U.S Army Research Laboratory and UT-Battelle, LLC, under Contract No. DE-AC0500OR22725 with the U.S. Department of Energy. The United States Government retains and the publisher, by accepting the article for publication, acknowledges that the United States Government retains a non-exclusive, paid-up, irrevocable, world-wide license to publish or reproduce the published form of this manuscript, or allow others to do so, for the United States Government purposes. The Department of Energy will provide public access to these results of federally sponsored research in accordance with the DOE Public Access Plan (http://energy.gov/downloads/doe-public-access-plan).

# 1. Introduction

Quantum information theory promises a variety of new techniques for transmitting, protecting, and processing information that may benefit the efficiency, security, and speed of tactical communication networks. This includes higher bandwidth encodings on network links, idealized encryption between network nodes, and faster, distributed computation across the network topology. These capabilities could offer disruptive advances in military communication networks. However, quantum information theory also imposes constraints on the operation of a quantum network including, for example, no-cloning and no-broadcasting. Therefore, conventional network engineering methods are unlikely to apply to the construction of future quantum networks, and new methods for quantum network engineering and management are needed to address different use cases. While multiple experimental demonstrations have validated these ideas, demonstrating the integration of quantum communication methods with existing protocols, channels, and data forwarding mechanisms is an open challenge.

Software-defined networking (SDN) is an emerging and fast growing technology for interconnecting network devices and forwarding packets based on unified policies and security enforcements. SDN capabilities have also been recently highlighted as an important enabling capability for military tactical networks, where assured networking is critical and flexible management are needed. The defining feature of SDN is deep programmability of the network at all layers including the extension of the network state into applications for enabling better pathing decisions. This versatile and programmable network architecture is expected to be more suitable to support future heterogeneous systems and implement ad hoc network policies, for example, that may arise in tactical environments. This includes quantum communication models that make use of quantum optical signals to encode and transmit information [5]. Previous models of quantum communication metadata have addressed software-defined communication [6], but there is still no information about how software-defined quantum metadata transport could be implemented. Building robust quantum network architectures and protocols is clearly a complex task that involves coordination between both the quantum and classical channels [7]. SDN principles may be able to manage the complex coordination required by future quantum networks.

In this paper, we address the inclusion of quantum metadata communication in the SDN paradigm by showing its compatibility with the OpenFlow protocol. The OpenFlow protocol plays a key role in enabling SDN architectures through its simple, flexible and adaptable programming interface [1]. OpenFlow provides a clean implementation of the data and control plane separation that is essential to the success of the SDN paradigm. Its external, centralized control plane enables programing of unified forwarding and security policies and supports the creation of flexible, reconfigurable networks as illustrated in Fig. 1. Although OpenFlow was originally designed for working with packet switched networks, later versions have included support for circuit switched networks that enable a converged control plane architecture for dynamically switching circuits and forwarding packets. OpenFlow has even proved to be more efficient than GMPLS label switching for managing optical networks [2].

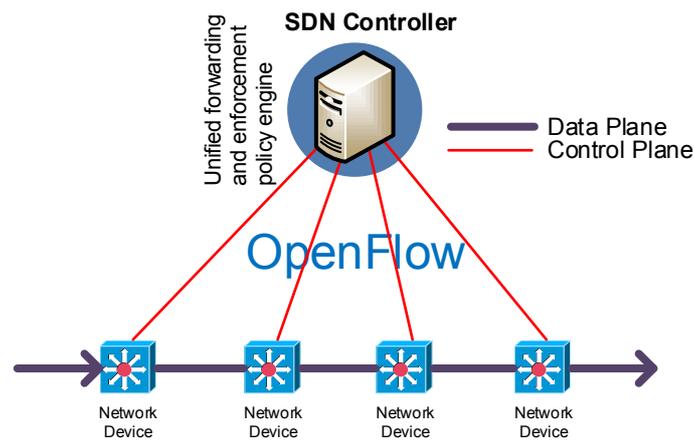

**Figure 1.** A schematic overview of SDN implemented with OpenFlow.

We investigate how quantum communication networks can be built using SDN principles and how the OpenFlow protocol can be extended to account for the metadata that is specific to quantum networks. Because of its programmability and compatibility with management of optical networks, OpenFlow is highly suitable to control the new attributes defining the classical channel that carries metadata between various quantum devices. In particular, OpenFlow version 1.4 extensively supports attributes specific to lambda switching in optical networks that we will show allows the ability to manage quantum optical channels interconnecting quantum network devices. However, development of network protocols and channel access specifications are needed to integrate quantum metadata communications within OpenFlow. We use OpenFlow extensions to transport quantum metadata attributes between quantum devices and the centralized OpenFlow controller. Quantum metadata extensions to OpenFlow enable attributes to be forwarded to other devices and applications in a quantum network by using a media agnostic classical channel on an optical network.

The paper is organized as follows. In Sec. 2, we provide an introduction to the principles of quantum communication including the unique concerns of encoding and protecting quantum states. In Sec. 3, we introduce a new networking layer, QPHY, to express the quantum physical layer used to transmit and receive quantum optical signals. In Sec. 4, we define the quantum metadata attributes expected for two different types of quantum network devices, i.e., quantum repeaters and quantum memories. In Sec. 5, we discuss how quantum metadata for these devices can be managed by the OpenFlow protocol and we provide specific prototype extensions for doing this. We present models for the link layer interaction between quantum devices that account for the role of the OpenFlow agents and controller in the network. These models provide implementations of the OpenFlow protocol that tested using a network controller. We then conclude in Sec. 6 by discussing additional extensions to OpenFlow needed to manage the quantum metadata attributes that are likely to arise in future quantum networks.

## 2. Overview of Quantum Communication and Quantum Networking

Quantum communication is distinguished from conventional (classical) communication by its use of quantum physical phenomena for encoding, transmitting, and decoding information [8]. In particular, photons, as quantum mechanical particles, may encode information that is not accessible in a classical context. This includes normalized superpositions of binary values, i.e., 0's and 1's, expressed as qubits, as well multi-photon entangled states that exhibit non-local correlations. With respect to communication, entanglement may be viewed as a resource that is generated and consumed during a protocol [9]. The presence of entanglement is necessary for the unique features of quantum communication, including teleportation, dense-coding, and secure key distribution. These quantum resources are prone to errors and noise that include decoherence, which is the loss of entanglement, as well as bit-flip and phase-flip errors in the qubit values. Overcoming these errors has led to the development of quantum error correction (QEC) codes and control techniques to either correct or circumvent information loss during noisy transmission environments [10, 11]. Thus, a quantum communication protocol operating on a realistic quantum channel must incur some resource overhead associated with error correction code and control mechanisms.

A notable use of entanglement is quantum teleportation, which permits the non-local exchange of a quantum state between two users. Quantum key distribution (QKD) is another example, in which the non-local correlations observed between photons in an entangled state can be used to generate a correlated sequence of bits. These bits then serve as a synchronized entropy source between two uses for purposes of encryption. Super-dense coding is a third example, in which entanglement is used to perform a unique function, in this case compression. Finally, distributed quantum communication is a general example in which multiple users coordinate their individual processing efforts while using entangled resources. Quantum networks may be used for distributed quantum computation or quantum sensing protocols. The distribution of entanglement between users plays a fundamental role in establishing the resources needed for quantum communication. In light of decoherence and photon loss mechanisms, there are practical limits on the distance over which an entangled photon state can be transmitted. Due to the quantum no-cloning principle, it is not possible to 'amplify' an entangled state or to 'repeat' the state as is typically done for extending the range of classical transmission. Instead, several schemes for extending transmission distance based on entanglement purification methods have been developed. These approaches distill multiple noisy copies of an entangled state into a single high fidelity resource. Concurrent storage of quantum states is challenging and the development of quantum repeater networks is currently a focus of intense research.

Numerous demonstrations of quantum communication systems have focused on proof-of-principle demonstrations limited to point-to-point communication. One of the first significant efforts at more robust quantum networking was by DARPA in 2002 [12] with more recent efforts in Europe [13], Japan [14], and China [15]. These efforts have addressed questions of how to integrate quantum communication within a network based on dedicated dark fiber, wavelength division multiplexing, or a mixture of quantum optical channels. However, all of these demonstrations have targeted application-specific designs, most notably, QKD. They represent fixed-point solutions and, in the broad sense of quantum network science, they have limited functionality. This has left open the question of how quantum networking can be made adaptive and developed in line with more modern networking approaches, such as SDN, as well as how to integrate these communication models with existing infrastructure.

Although quantum network development has received significantly less attention than nodal and link engineering, there is still pressing research in the field of quantum network science [19-22]. Quantum network science investigates the capabilities that quantum information carriers add to network based applications [23, 24]. More directly related to the development of quantum network architecture is the work of van Meter and Touch, who have discussed the relevance of separating concerns in the design of a quantum repeater network [25, 26]. Those results support our claim that the protocol layers required for quantum networking must separately account for functional concerns. In the case of a quantum repeater network, van Meter and Touch provide a protocol stack that includes purification and error correction operations as separate layers. Van Meter, Touch, and Horsman have elaborated on the design of recursive quantum repeater networks [27] that can be used for achieving scalability in the repeater networks.

Humble and Sadlier have previously suggested a possible software defined transport model for transmitting the metadata [6]. That work was limited in context to a single quantum network application, namely classical communication using super-dense coding. The quantum physical layer was based on polarization-encoded qubits, while the control and data layers were software defined. This implementation of the super-dense coding protocol permitted a sender Alice to transmit a classical message to the receiver Bob. The software-defined data layer was then used to implement the encoding and decoding procedures for the transmission, including classical error correction [28]. We extend this previously postulated communication model to accommodate SDN functionality and transport quantum metadata across the network and make it available to other components through an SDN controller. This requires the middleware within a quantum network device (node) to interface with the underlying quantum hardware that receives quantum optical transmissions. The network device must perform quantum data processing, for example, by measuring single photon state and recoding their arrival times to classify any associated quantum state attributes.

## 3. Quantum Physical Sublayer (QPHY)

Our design of SDN for quantum communications begins by decomposing the capabilities provided by quantum protocols into the traditional networking layers [5, 6]. These layers however must be extended to accommodate the unique behaviors required for quantum networking. This includes entanglement distribution, entanglement purification, and quantum error correction. We define a quantum physical layer within this stack to expresses quantum channel access. Abbreviated here as QPHY, the quantum physical layer has been realized previously in state of the art network testbeds but its role in the networking stack has not been explicitly defined. We develop an interface for the QPHY layer through which metadata defining the quantum physical behavior is passed. The QPHY interface permits management of the quantum network using the principles of SDN. In particular, the generation of entanglement within a quantum network arises from coordination of the classical and quantum control planes defining the network.

A typical quantum network of interacting devices, such as transmitters, receivers, repeaters and routers, may be decomposed into major functional layers based on separations of concerns. As with conventional networks, quantum networks require management, control, data, and physical layers [5]. Links between devices establish quantum communication channels that are then used for transmitting quantum states of information. Network nodes must also make use of classical communication channels to carry information characterizing the usage of the quantum channel, leading to the definition of quantum metadata [6]. More broadly, quantum metadata specifies the attributes needed by a network to accommodate applications requesting access to the quantum communication channel.

The QPHY layer is realized by the network device responsible for encoding and decoding, transmitting and receiving, repeating and routing quantum information. We focus exclusively on the use of quantum optics for expressing QPHY. Quantum optical technology shares many of the same basic operating principles and components as conventional optical communication. This includes pulsed transmissions at optical wavelengths over fiber. However, there are also several important distinctions between quantum and conventional optics. Foremost is the extremely low power used in quantum optical system, which effectively have an average photon number of one or less. In addition, information may be encoded in time, polarization, orbital angular momentum or combinations of these degrees of freedom, in addition to the conventional quadrature variable.

Furthermore, there are two variants of quantum channels for consideration, namely, discrete and continuous variable optical modes. Both types are capable of supporting quantum information but they differ significantly in the physical encoding. A typical example of a discrete variable channel uses the orthogonal states of photonic polarization to encode binary values. A continuous variable channel performs the same task, for example, using quadrature variables of a weak optical field. In both cases, these encodings support coherence and entanglement between channel uses due to quantum physical phenomena. We will not address further the distinctions of quantum topical technology, as our SDN-based approach to quantum network management intends to be flexible with respect to the physical encoding.

Functionally, QPHY represents the translation of classically encoded information into a quantum encoding, or vice versa. For a quantum transceiver, this means mapping classical input symbols to elements of an alphabet of quantum states or, conversely, projecting a quantum state via measurement into a classical value. For a quantum memory or repeater device, QPHY must implement operations that include storing intermediate quantum states and performing quantum logic across multiple channel uses. A typical example is an entanglement distillation protocol used to overcome decoherence by transforming many lossy channels into a single high-fidelity channel. This requires parsing the purification protocol into the operational sequence and reporting the successful channel creation back to the upper layers. More generally, the functionality of the underlying quantum optical hardware determines the diversity of the QPHY interface. The ability to standardize a common interface to the QPHY layer is needed for supporting the extension of SDN to reconfigurable quantum networks.

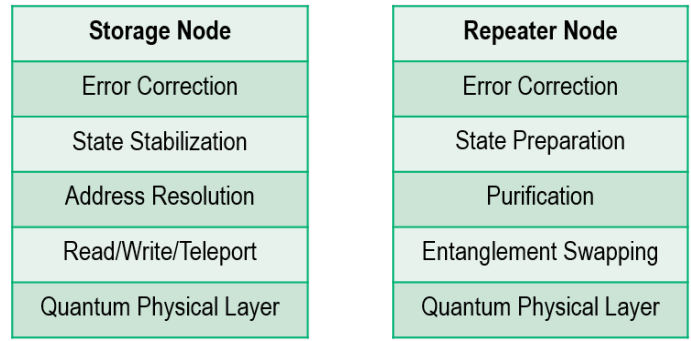

**Figure 2**. A representation of the logical operations performed by two quantum network devices (left) a quantum storage node and (right) a quantum repeater node.

As an example of the functional requirements for quantum network devices, consider a network composed from quantum memory and repeater nodes. Representations of the data layers expressed by these nodes are shown in Fig. 2. A quantum memory node stores quantum information that may be acted upon remotely by a read or write request or locally by a measurement request. In between requests, the memory node maintains the coherency of the stored information. This is accomplished using a quantum error correction (QEC) protocol, which protects against information loss to the surrounding environment. The implementation of a QEC code must be tailored to the physical layer of the node as well as the period of time for which information is stored. In addition, a storage node may apply a control technique, such as state stabilization, to suppress environmental noise. These protocols must be executed using nodal resources that are resolved to particular storage locations, i.e., network addresses. Intra-node address resolution is also required for carrying out read/write requests from other nodes. By comparison, a quantum repeater node links quantum communication channels between remote devices by performing entanglement swapping or teleportation operations. The repeater node may also use QEC and control protocols for improving the quality of service of the quantum channel, i.e., for increasing its entanglement. A repeater node uses state preparation methods that apply purification and entanglement swapping stages. Each stage is responsible for preparing the resources needed to ensure transmission quality. Our example has differentiated between the storage and repeater operations. Of course, an actual network node may support both storage and repeater functionality. In this case, the design of the node will determine how these functionalities are merged, e.g., as single, mixed or independent interfaces.

As indicated by the description of the storage and repeater nodes, the different behaviors of a quantum network device can be controlled within a software-defined networking environment. For both the storage and repeater nodes, the QEC and control protocols represent metadata information that tunes functionality. Read/write requests will be part of the conventional data message, while channel access specifications, including channel wavelength and routing information will be part of the control plane.

Finally, as a point of comparison, the physical layer in a conventional optical network, abbreviated as PHY, expresses the optical communication channel including hardware for encoding and transmitting classical states of information. The PHY layer is also essential for the operation of a quantum network, where it may be used for standalone and handshaking protocols, or used to carry side-channel information for quantum transmissions. Standards and interfaces for conventional PHY layers are well defined for a variety of transmission medium including optical communication, e.g., IEEE 802.3 standards. The overall communication protocol specifying how users exchange classical information, for example, to carry out an application specific task, can be implemented using conventional networking over the PHY layer. In particular, TCP and other Internet protocols are largely sufficient as is for exchanging these types of classical messages.

## 4. Quantum Metadata Specification

Quantum metadata specifies the attributes needed by QPHY to accommodate specific uses of the quantum channel. This is classical information that moves through the network and characterizing how applications intend to make of the quantum network. Each type of network device may have different metadata requirements, e.g., a router versus a transmitter, but the metadata should be consistent across devices of the same type. In Table 1, we present a prototype specification of the quantum metadata needed to manage the behavior of quantum network storage and repeater devices. These specific metadata fields include identifiers and specification for the communication protocol, the error correction protocol, and the quantum channel access. For example, the QCHANNEL field is a provided as a unique identifier that specifies the quantum channel to which the metadata applies. This would include quantum network address information that signifies the source and destination of the quantum signal. The QCHANNEL_SPEC field specifies additional parameters used for characterizing the transmission and reception of the quantum signals. This may include wavelength, power, symbol rate, etc. A quantum network device must be capable of supporting transmission of the requested specification in order for communication to be successful.

**Table 1.** Quantum Metadata Specification.

| Metadata Field | Description |
| --- | --- |
| **QCHANNEL** | *Unique identifier for channel on quantum network* |
| **QCHANNEL_SPEC** | *QCHANNEL parameters for TX and RX* |
| **QCOM** | *Identifier for information theoretic protocol* |
| **QCOM_SPEC** | *Parameters for QCOM field* |
| **QEC** | *Quantum error correction protocol identifier* |
| **QEC_SPEC** | *Quantum error correction protocol specification* |

In Table 1, the QCOM field identifies the type of protocol to be implemented between devices. Example tasks include QKD, quantum teleportation, and dense coding. The QCOM_SPEC field identifies additional specifications for how the protocol should be implemented. This may include specifying the protocol needed for QKD, the remote-state and gate preparation steps for quantum teleportation (QT), and the encoding and decoding tables used for super-dense coding (SDC). For the repeater and storage nodes, the QEC field identifies the quantum error correction protocol that should be used for detecting and correcting errors in the hardware. Examples include specific error correction strategies. The QEC_SPEC field would specify the encoding and decoding methods as well as verification circuits to be used during the correction procedure. These indicated methods and circuits implicitly impose requirements on how and when the quantum channel is access but the metadata does not define how the physical layer should achieve this implementation.

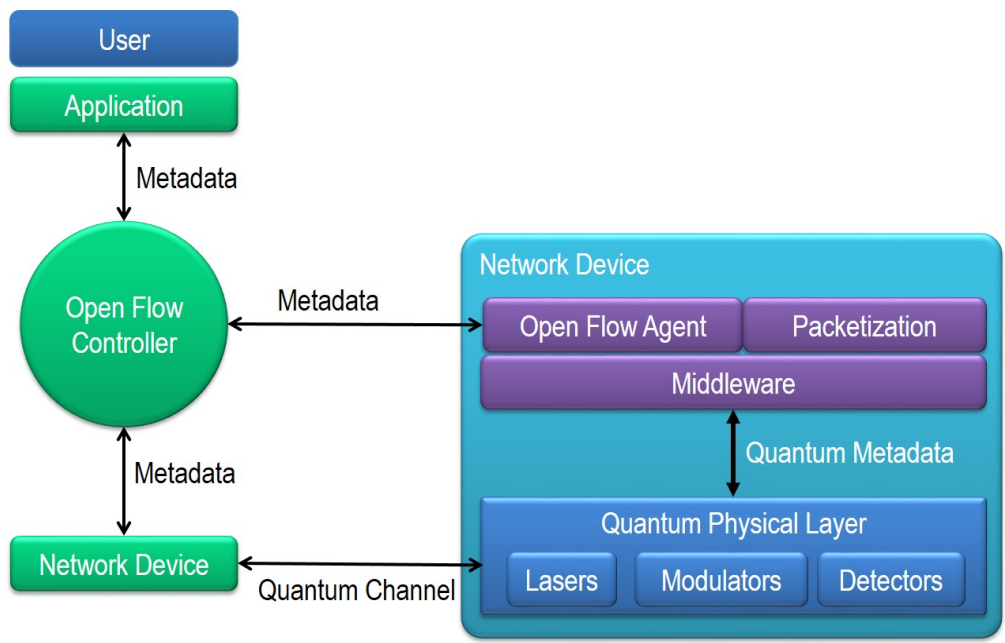

**Figure 3.** A schematic of the flow of metadata between an OpenFlow-enabled quantum network device and an OpenFlow controller managing network protocols.

## 5. OpenFlow Quantum Metadata Communication Model

OpenFlow is a communications interface defined between the control and forwarding layers of an SDN architecture. Its specification allows direct access and manipulation of the forwarding plane of network devices such as switches and routers. In a conventional network, the functionality of a switch is controlled by its own local software. SDN and OpenFlow in particular, centralize the decisions about how to specify functionality. These permits the entire network to be programmed independently of the individual switches. We extend the use of the OpenFlow protocol to accommodate the quantum metadata necessary for specifying quantum communication between network nodes. In particular, the OpenFlow protocol is rapidly developing to provide additional functionality. OpenFlow version 1.3 offers IPv6 support [3], while optical extensions are one significant feature added in version 1.4. In particular, optical port properties added in OpenFlow 1.4 offers opportunities to configure and monitor optical frequencies and power of transmitter lasers [4]. This new feature is a key element in bringing the SDN framework closer to optical network management and programming. In addition, optical network management functions enable OpenFlow to compete with 3-way handshake (3WHS) protocols for providing dynamic path discovery and path setup for optical networks. OpenFlow also supports topology discovery and network instrumentation for measuring performance statistics. These qualities have made OpenFlow one of the single most important standards-based protocols driving SDN architectures today. In particular, OpenFlow makes it easy to create network architectures that interconnect new types of network devices.

Given a quantum metadata specification, a quantum network must manage the flow of this metadata between network devices to support application requests. This can be accomplished using the OpenFlow specification. Extending the OpenFlow layer of a quantum network device requires that a new OpenFlow table be defined to hold quantum metadata metrics. This table, denoted as QCM, must hold the current quantum metadata state of the device and may be overwritten by the local middleware due to changes in the underlying quantum communication channel (Fig. 5). The QCM table is queried by the OpenFlow controller for poling the QCM state of each device within a given topology to establish a unified view of each device QCM under its control. From the controller, other devices and applications can learn about the QCM state of any of all of the devices managed by the controller. In this model, the metadata fields for each node are polled by the OpenFlow controller and, when necessary, are changed by the node when a request from the OpenFlow controller is received.

Metadata generated by the network device is converted to OpenFlow flags by a flow module within the OpenFlow agent. A modified OpenFlow controller polls and collects these quantum metadata attributes from those devices capable of providing such data. The flow module extracts metadata from the middleware and transforms these attributes into the corresponding OpenFlow statistics (Fig. 4). Extensions to the OpenFlow statistics tables are written to accommodate the new set of statistics specific to quantum metadata. The OpenFlow flags are handed over to OpenFlow daemon for passing the metadata in the OpenFlow format to the controller.

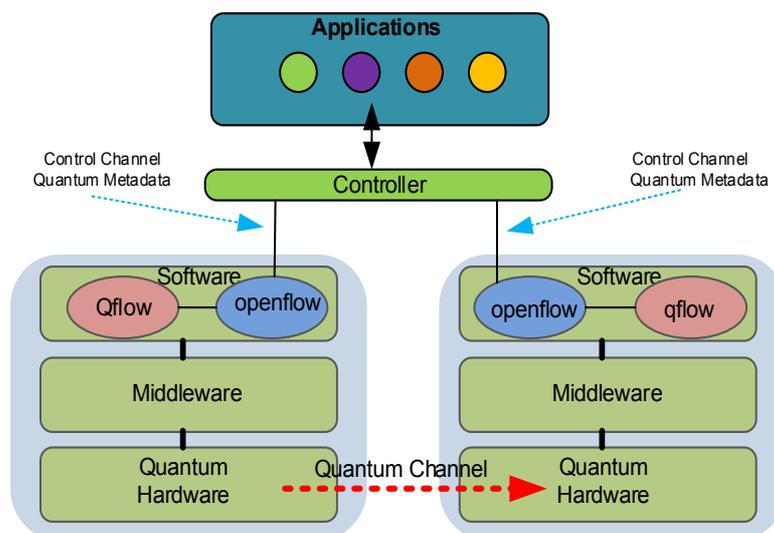

**Figure 4.** Proposed Quantum Network Device Model with OpenFlow Control Channel

Quantum metadata can be encoded as part of the OpenFlow statistics and information between the device and the controller can be exchanged synchronously via a set of request and response directives. Quantum metadata can also be transferred to the controller asynchronously from the device whenever the local metadata counters are changed or modified by the device middleware. In OpenFlow, state information and statistics are exchanged using multi-part messaging directives to accommodate information need to break down larger messages into multiple messages, each not exceeding the size of 64Kb. Metadata for quantum communication can follow the multipart messaging format to make sure all the attributes are accommodated.

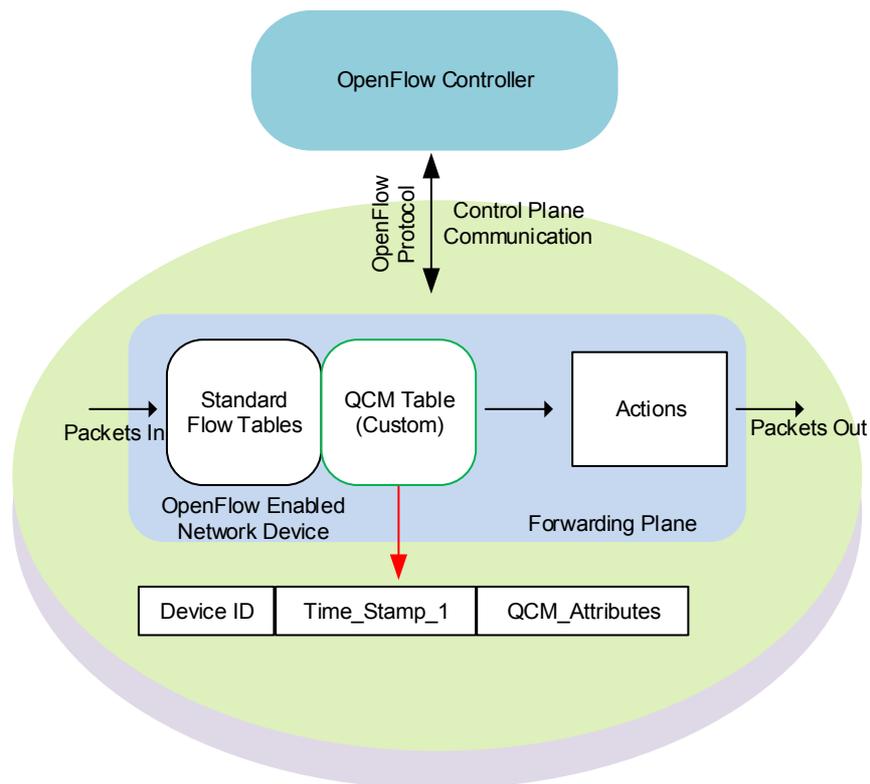

**Figure 5.** A new OpenFlow table to accommodate metadata attributes

The controller can establish QCM synchronization either by synchronous or asynchronous methods that are defined in the OpenFlow specification. Synchronous communications are a set of request and reply messages to exchange the QCM table state between the quantum device and the controller using a multi-part message format. Asynchronous methods are used when the QCM state changes and a device needs to update the controller without waiting for an upstream request. Asynchronous updates will propagate the new QCM data as it arrives without any delay.

The default request/response message primitives in OpenFlow are sufficient to accomplish synchronization for quantum network devices. A new OpenFlow table called OFPMP_QCM is created to hold all the metadata attributes. However, it is necessary to modify the response messages to include the QCM table states. In particular, the request headers of the multi-part message response must be changed to support quantum metadata attributes. As shown in Figs. 6 and 7, new attributes are stored in the QCM group table for processing and applying necessary actions to the incoming traffic. The newly defined OpenFlow table will be added to the pipeline for processing the traffic hitting the controller during the packet-in events. If there is a match on any of the QCM table attributes, the programmed actions will be applied to the flows.

| Request-Header | Request-Body |
|---|---|
| `struct ofp_qcm_request{` | `struct ofp_qcm_stats{` |
| `struct ofp_header header;` | `uint16_t QCHANNEL` |
| `uint16_T type;` | `uint16_t QCHANNEL_SPEC` |
| `uint8_t pads[4];` | `uint16_t QCOM` |
| `uint16_t flags;` | `uint16_t QCOM_SPEC` |
| `uint8_t body[];` | `uint16_t QEC` |
| `};` | `uint16_t QEC_SPEC` |
| `OFP_ASSERT(sizeof(struct ofp_multipart_request) == 16);` | `};` |
| `enum ofp_multipart_request_flags{` | `OFP_ASSERT(sizeof(struct ofp_QCM) == 56);` |
| `OFPMPF_REQ_MORE = 1 << 0};` | |

**Figure 6.** Modified OpenFlow multi-part message request header and body

| Response-Header | Response-Body |
|---|---|
| `struct ofp_qcm_reply{` | `struct ofp_qcm_stats{` |
| `struct ofp_header header;` | `uint16_t QCHANNEL` |
| `uint16_T type;` | `uint16_t QCHANNEL_SPEC` |
| `uint8_t pad[4];` | `uint16_t QCOM` |
| `uint16_t flags;` | `uint16_t QCOM_SPEC` |
| `uint8_t body[];` | `uint16_t QEC` |
| `};` | `uint16_t QEC_SPEC` |
| `OFP_ASSERT(sizeof(struct ofp_multipart_reply) == 16);` | `};` |
| `enum ofp_multipart_reply_flags{` | `OFP_ASSERT(sizeof(struct ofp_QCM) == 56);` |
| `OFPMPF_REQ_MORE = 1 << 0 /* more replies to follow */` | |
| `};` | |
| `ofp_qcm_multipart_request` | |
| `OFPMP_QCM = 17` | |

**Figure 7**. Modified OpenFlow Multi-part message response header and body

We have used the quantum metadata attributes to define an SDN model for a classical control channel to transport, centrally store and distribute the quantum metadata in SDN enabled network topology

13containing quantum devices. In order to verify the functionality of the control channel, a simple simulation of the proposed channel is created using python based discrete event simulation framework which is suited for rapid simulation of proposed protocol operations and network states. Quantum metadata attributes exchange is simulated by writing a custom python class called QDC that simulates communication between the SDN controller and the quantum network device. Each of the attributes exchanged between the controller and the quantum devices is printed out to the standard output as confirmation about the state of the device.

We currently use these simulations as a fundamental demonstration of how the quantum device can be integrated with the SDN controller. While our efforts to model the underlying protocols are ongoing [29], we have used this functional model to verify the protocol syntax. In particular, every time an attribute is changed or updated, the OpenFlow agent on the device updates the SDN controller with new metadata that is in turn made available to all other quantum devices under its control as needed. It is also possible and makes sense to build an authentication process before the device is allowed to read from and write to the controller to ensure security. The goal of this simulation is to calibrate the timing of the channel and flow of the data exchange between the controller and the nodes before the actual implementation of the metadata channel in real network topology using modified OpenFlow with the proposed metadata data structures. Since the timing of control information flow in control channel is critical for the programmable switches, controller in the path to properly map classical channel connection intents expressed in metadata attributes with the associated quantum channel routing correctly. We have used a simple discrete event simulation framework called SimPy to model the request and response mechanisms of OpenFlow protocol. As shown in Fig. 8, the SimPy model emulates the behavior of messaging between the agent and the controller.

```
Quantum METADATA STATE change sensed. Collecting QMD attributes at 0
QMD Requested by Controller............
Receiving QMD attributes ..... begins:
QPROTO: Binary Dense Coding  Received
QPROTO_SPEC: ####### Received
QCHANNEL: ####### Received
QCHANNEL-SPEC: ###### Received
QPROTO_SPEC: ###### Received
Waiting for STATE change
------------------------------------------
Quantum METADATA STATE change sensed. Collecting QMD attributes at 5
QMD Requested by Controller............
Receiving QMD attributes ..... begins:
QPROTO: Binary Dense Coding  Received
QPROTO_SPEC: ####### Received
QCHANNEL: ####### Received
QCHANNEL-SPEC: ###### Received
QPROTO_SPEC: ###### Received
Waiting for STATE change
------------------------------------------
Quantum METADATA STATE change sensed. Collecting QMD attributes at 10
QMD Requested by Controller............
Receiving QMD attributes ..... begins:
QPROTO: Binary Dense Coding  Received
QPROTO_SPEC: ####### Received
QCHANNEL: ####### Received
QCHANNEL-SPEC: ###### Received
QPROTO_SPEC: ###### Received
Waiting for STATE change
------------------------------------------
15.0
```

**Figure 8**: OpenFlow based quantum metadata channel simulation output

## 6. Conclusions

SDN provides a flexible approach to managing a network of diverse devices, and its continued evolution highlights an opportunity to include emerging technologies. We have taken advantage of this flexibility to show that the OpenFlow protocol is capable of extending SDN principles to the emerging field of quantum networking. We have described a broad set of use cases for quantum networks and defined how the QPHY layer needed for quantum communication channels can be integrated with the management of existing network layers. We provided a prototype specification for the types of quantum metadata expected to be shared between network devices used to store and forward quantum information, i.e., quantum memories and repeaters. We have also developed the OpenFlow specifications need for the controller to request updates from devices and the responses those devices would send. The current specification offers management of mid-level device behaviors, including communication encoding and error correction methods as well as transmission specifications such as channel frequency.

Within the OpenFlow paradigm, a quantum network device exchanges metadata with the network through the OpenFlow controller. This permits polling of the quantum network using the OpenFlow controller and the extension to the flow tables. We have presented a prototype specification for how this metadata exchange would define attributes of quantum network devices. We have also described modifications for how the OpenFlow framework can permit other network devices to share this quantum metadata, where our specification has explicitly taken advantages of the extensible nature of the OpenFlow specification. In particular, we have used OpenFlow to generate a programmable network that can manage the transport of quantum metadata. Although our initial demonstration has focused on very simple aspects of quantum networking, i.e., managing protocols between devices, it should be apparent that the extensible nature of the specification provides a broad applicable framework. This work establishes how to leverage SDN principles within quantum networks by integrating quantum metadata with existing management methods.

OpenFlow provides the necessary programmable network interface for creating quantum communication channels and with the recent inclusion of SDON management features in OpenFlow version 1.4 this programming may make use of existing optical devices. Our simulation shows that OpenFlow is viable for quantum metadata transport via classical channels. By leveraging the principles of SDN, we believe this framework can also extend benefits to applications using northbound API. Future integration with applications, such as super-dense coding and QKD for efficient and secure messaging, offer the possibility to accelerate deployment of novel quantum communication methods. The OpenFlow implementation we have developed eases the acceptance of new hardware and algorithms into existing and future networks.

Future uses of OpenFlow for quantum networking will require tracking of additional metadata, most notably, metadata to define the state of quantum switches and routers [29]. Quantum switches represent a component of the quantum network responsible for connecting devices over established input and output paths. Because quantum optical states cannot be copied or probed during transmission, management of network switching and routing must occur by sharing metadata across the network. This includes network discovery of the quantum devices as well as change in states of the switch itself. The OpenFlow protocol explicitly supports this type of metadata sharing, but the procedures by which the network informs both the conventional and quantum switches has yet to be determined. Future studies of quantum metadata are needed to identify the necessary extensions to the OpenFlow tables.

**Acknowledgments**

The authors thank the US Army Research Laboratory for its support in carrying out this work. VRD expresses his gratitude to Applied Research for the Advancement of S&T Priorities (ARAP) program for its partial financial support. TSH acknowledges support from the Defense Treat Reduction Agency. The authors thank Dr. Vinod Mishra for participating in the discussions during the initial stage of this project.

**Author Contributions**

VD and TSH contributed equally to the development of research results and the preparation of the manuscript.

**Conflicts of Interest**

The authors declare no conflict of interest.

**Author Biographies**

**Venkat R. Dasari** is a scientist at the Advanced Computing Architectures Division, U.S. Army Research Laboratory, Aberdeen Proving Ground, Maryland. Previously, he worked both in the Government and private sector architecting and programming networks. His current interests are focused on programmable classical and quantum networks, GENI Testbed Technologies and distributed mobile network Intelligence.

**Travis S. Humble** is director of the Quantum Computing Institute at Oak Ridge National Laboratory, Oak Ridge, Tennessee. He was previously an Intelligence Community Postdoctoral Research Fellow and is currently a Department of Energy Early Career Research scientist. His current research interest is the design and development of quantum computing technologies for next generation high-performance computing systems.